\documentstyle[preprint,aps]{revtex}

\draft

\begin{document}
\title{Neutral-dangling bond depletion in {\it a-}SiN films caused by magnetic
rare-earth elements}
\author{M. S. Sercheli and C. Rettori}
\address{Instituto de F\'{i}sica ``Gleb Wataghin'', UNICAMP, 13083-970, Campinas-SP,\\
Brazil}
\author{A. R. Zanatta}
\address{Instituto de F\'{i}sica de S\~{a}o Carlos, USP, 13560-250, S\~{a}o Carlos-SP,%
\\
Brazil}
\maketitle

\begin{abstract}
Amorphous silicon-nitrogen thin films doped with rare-earth elements ({\it a}%
-SiN:RE; RE = Y, La, Pr, Nd, Sm, Gd, Tb, Dy, Ho, Er, Yb, and Lu) have been
prepared by co-sputtering and studied by means of electron spin resonance
(ESR). It was found that the neutral dangling-bond density [D$^{0}$] of {\it %
a}-SiN films decreases with the presence of magnetic REs and the drop of [D$%
^{0}$] approximately scales with the spin and/or the de Gennes factor of
each rare-earth element. These results suggest that a strong exchange-like
interaction, ${\cal H}=J_{RE-D^{0}}{\bf S}_{RE}{\bf .S}_{D^{0}}$, between
the spin of the magnetic REs and D$^{0}$ may be responsible for this
behaviour, similarly to the decrease of $T_{c}$ in RE-doped superconductors.
\end{abstract}

\pacs{75.70.-i, 76.30.Kg, 78.66.Jg}

\section{Introduction}

Taking into account the recent technological advances, and needs of our
modern society, the study of the magnetic properties of new materials is of
fundamental importance to develop devices for different applications such
as, for example, memory structures.\cite{gritsenko} In view of their
characteristics, amorphous silicon ({\it a}-Si) thin films seems to be good
candidates for such a purpose.\cite{Street} An interesting way of studying
the {\it a}-Si thin films magnetic response is to focus on the properties of
the neutral dangling-bonds (D$^{0}$) present in these materials. Neutral
dangling-bonds are paramagnetic centers that are excellent probes for the
investigation of {\it a}-Si thin films.\ Moreover, silicon dangling-bonds
are charge trapping centers\cite{warren} that are more stable under the
diamagnetic D$^{+,-}$ form.\cite{robertson} In the present work we have
studied the behaviour of the paramagnetic defects D$^{0}$ in amorphous
silicon nitride thin films doped with various rare-earth elements, {\it a}%
-SiN:RE (RE = rare-earths: Y, La, Pr, Nd, Sm, Gd, Tb, Dy, Ho, Er, Yb, and
Lu). Depending on the RE dopant, these thin films present a relative strong
and narrow light emission, even at room temperature.\cite{Zanatta1},\cite
{Zanatta2} As a consequence, RE-doped {\it a}-SiN thin films\ are expected
to be ideal candidates to develop photonic devices. Towards this end, the
study of their magnetic properties will certainly contribute to decide about
the potential applications of these materials.

\section{Experimental}

All films were prepared in a high vacuum chamber (base pressure $\sim
2x10^{-6}$ Torr), by radio frequency ($13.56$ MHz) sputtering a Si ($99.999$
\% pure) target covered at random with small pieces of metallic RE ($99.9$
\% pure) elements. Polished crystalline ({\it c}-)Si wafers and high-purity
quartz plates were used as substrates in every deposition run. During
deposition, the substrates were kept at $\sim 70$ 
${{}^o}$%
C under a constant total pressure of $\sim 5x10^{-3}$ Torr consisting of a
mixture of high-purity Ar + N$_{2}$ gases. The mean thickness of the films
was $\sim 500$ nm.

The atomic composition of the {\it a}-SiN:RE films ($\sim 40\%$ Si, $\sim
59\%$ N, $\sim 0.6\%$ RE) were determined by {\it Rutherford} backscattering
spectrometry (RBS) in the case of Si and RE and by nuclear reaction analysis
(NRA) for N. A non-intentionally amount of hydrogen of $\sim 1-2\%$ H \ was
detected by elastic recoil detection (ERD) analysis in all {\it a-}SiN:RE
films. The density of\ the films was estimated to be $\sim 8x10^{22}$ at. cm$%
^{-3}$. The optical bandgap of these films were determined through optical
transmission measurements in the visible-ultraviolet range using a
commercial spectrophotometer and stays around $5.5$ eV.\cite{Zanatta2} Room-$%
T$ {\it Raman} scattering measurements, using the $488.0$ nm line of an Ar$%
^{+}$ {\it laser}, were also performed and confirmed the amorphous structure
of the films.

The electron spin resonance (ESR) experiments were carried out in\ a {\it %
Bruker} X-band ($9.47$ GHz) spectrometer using a room-$T$ TE$_{102}$ cavity.
All mesuarements have been taken at room temperature.

\section{Results and Discussion}

This work presents a new approach in the study of the D$^{0}$ density of 
{\it a}-SiN thin films doped with REs. Our main finding was the depletion of
the density of D$^{0}$ in the {\it a}-SiN matrix caused by the presence of
magnetic RE species. We have observed that the insertion of magnetic RE
species dramatically suppresses the number of ESR active D$^{0}$ states and
that such a decrease approximately scales with the spin component of the RE
magnetic moment.

Table I displays the atomic concentrations [RE], [Si], [N] and [H] as
determined from RBS, NRA, and ERD for all the films investigated in this
work. From the D$^{0}$ ESR intensity measurements, and using as standard a
KCl-pitch sample, we have estimated the [D$^{0}$] of each film. The D$^{0}$
ESR parameters and [D$^{0}$] are also given in Table I. As can be seen from
Table I, the average density of D$^{0}$ magnetic defects in these films was
of, typically, $\sim 2x10^{20}$\ cm$^{-3}$.

Figure $1$ shows the room-$T$ ESR normalized spectra of D$^{0}$ in {\it a}%
-SiN films doped with different RE elements (notice the different
intensities). From a Lorentzian lineshape fitting of the resonances for all
the {\it a}-SiN:RE thin films we have obtained approximately the same
peak-to-peak linewidth, $\Delta H_{pp}\simeq $ $18(2)$ G, and field for
resonance, $H_{R}\simeq 3378(1)$ G (corresponding to $g\simeq 2.004(1)$). An
early ESR study on undoped {\it a-}Si$_{3}$N$_{4}$\ films pointed out a very
weak D$^{0}$ ESR signal with $g=2.0054$\ and $\Delta H_{pp}=3.9$\ G.\cite
{shames} A comparison of these data with our larger linewidth and higher [D$%
^{0}$] suggests that the present {\it a-}SiN:RE films are more disordered.
This is probably associated to the higher N/Si ratio ($\geq 1.4)$\ in our
films.\cite{Bandet}

Figure $2$ shows [D$^{0}$] for the various RE elements investigated in this
work. It is noted that the magnetic REs cause a dramatic depletion of [D$%
^{0} $] and the strongest suppressing effect is found for Gd$^{3+}$, at the
middle of the RE-series. Within our experimental accuracy, the non-magnetic
RE elements do not cause a systematic change in [D$^{0}$]. The inset of Fig. 
$2$ presents the drop of [D$^{0}$], or in another words, the number of
inactive ESR D$^{0}$, 
\mbox{$<$}%
[D$^{0}$(RE$_{nm}$)]%
\mbox{$>$}%
$-$[D$^{0}$(RE$_{m}$)], due to the presence of the magnetic RE$_{m}$'s
relative to the average value for the non-magnetic RE$_{nm}$'s. Notice that
the minimum in Fig. $2$ correlates quite well with the RE's de Gennes
factor, $(g_{J}-1)^{2}J(J+1)$, and/or the $S(S+1)$ factor.

The striking result of Figure $2$ suggests that the mechanism responsible
for the depletion of [D$^{0}$] involves the spin part of the RE magnetic
moment and may be attributed to a strong exchange-like coupling, ${\cal H}%
\sim J_{RE-D^{0}}{\bf S}_{RE}{\bf .S}_{D^{0}}$, between the RE$^{3+}$ spin, $%
{\bf S}_{RE}$, and the spin of the D$^{0}$, ${\bf S}_{D^{0}}$. Such a strong
exchange coupling may probably shift and/or broaden the D$^{0}$ resonance
beyond the detection limit of our ESR experimental facilities. It is then
possible that a coupling of this kind leads to a [D$^{0}$] decrease
involving the de Gennes factor, $(g_{J}-1)^{2}J(J+1)$. The existence of this
factor has been largely confirmed in RE-doped type II superconductor
compounds through the decrease of the superconducting temperature, $T_{c}$, (%
$\Delta T_{c}/\Delta c<0$) due to the cooper-pairs breaking property of the
RE ions. \cite{Maple} \cite{Gschneider} \cite{Abrikosov} At this point, we
should also mention that the depletion of [D$^{0}$] could be approximately
described by the spin part of the RE-magnetic moment, $S(S+1)$, that also
takes its highest value at the Gd$^{3+}$ ion ($J=S=7/2$). These two analyses
are showed in the inset of Fig. $2$.

\section{Conclusions}

Rare-earth doped amorphous silicon-nitrogen films were prepared by
co-sputtering and investigated by means of different experimental techniques
(ESR, $dc$-susceptibility, ion-beam analyses, and {\it Raman} scattering).
The main experimental results can be summarized as: $i)$ a strong depletion
in the density of D$^{0}$ states induced by the presence of magnetic RE\
ions, and $ii)$ the correspondence between this depletion and the RE's de
Gennes factor, $(g_{J}-1)^{2}J(J+1)$, and/or the RE's $S(S+1)$ factor. These
results led us to propose a mechanism involving a strong exchange-like
coupling between the RE$^{3+}$ magnetic moment and the spin of the silicon
dangling-bond D$^{0}$. This strong coupling may cause a large shift and/or
broadening of the D$^{0}$ resonance which are beyond the limit of detection
of our ESR spectrometer.

\section{Acknowledgments}

This work has been supported by FAPESP, CAPES and CNPq.

$
\begin{tabular}{cccccccc}
Film & [Si]$_{\text{RBS}}$ & [N]$_{\text{NRA}}$ & [RE]$_{\text{RBS}}$ & $%
\Delta H_{pp}$ & $H_{R}$ & $g$ & [D$^{0}$] \\ 
& (at.\%) & (at.\%) & (at.\%) & (G) & (G) & - & (cm$^{-3}$) \\ 
$a$-SiN & $40.0$ & $58.0$ & $0.0$ & $16(2)$ & $3378(2)$ & $2.004(1)$ & $%
3.2\times 10^{20}$ \\ 
$a$-SiN:Y & $40.0$ & $58.0$ & $0.7$ & $20(2)$ & $3378(2)$ & $2.004(1)$ & $%
4.2\times 10^{20}$ \\ 
$a$-SiN:La & $41.0$ & $57.6$ & $0.4$ & $18(2)$ & $3378(2)$ & $2.005(1)$ & $%
2.7\times 10^{20}$ \\ 
$a$-SiN:Pr & $40.0$ & $58.0$ & $0.6$ & $16(2)$ & $3379(2)$ & $2.004(1)$ & $%
1.9\times 10^{20}$ \\ 
$a$-SiN:Nd & $39.0$ & $59.7$ & $0.8$ & $16(2)$ & $3378(2)$ & $2.005(1)$ & $%
9.6\times 10^{19}$ \\ 
$a$-SiN:Sm & $38.0$ & $59.2$ & $0.8$ & $15(2)$ & $3378(2)$ & $2.004(1)$ & $%
9.2\times 10^{19}$ \\ 
$a$-SiN:Gd & $39.0$ & $59.3$ & $0.7$ & $n.a.$ & $n.a.$ & $n.a.$ & * $%
2.5\times 10^{17}$ \\ 
$a$-SiN:Tb & $39.0$ & $59.3$ & $0.7$ & $19(2)$ & $3377(2)$ & $2.005(1)$ & $%
1.6\times 10^{20}$ \\ 
$a$-SiN:Dy & $38.0$ & $59.6$ & $0.4$ & $17(2)$ & $3377(2)$ & $2.005(1)$ & $%
1.7\times 10^{20}$ \\ 
$a$-SiN:Ho & $40.0$ & $58.0$ & $0.6$ & $16(2)$ & $3379(2)$ & $2.004(1)$ & $%
1.7\times 10^{20}$ \\ 
$a$-SiN:Er & $38.0$ & $59.5$ & $0.5$ & $16(2)$ & $3379(2)$ & $2.004(1)$ & $%
1.7\times 10^{20}$ \\ 
$a$-SiN:Yb & $39.0$ & $59.4$ & $0.6$ & $21(2)$ & $3378(2)$ & $2.004(1)$ & $%
2.4\times 10^{20}$ \\ 
$a$-SiN:Lu & $41.0$ & $57.7$ & $0.3$ & $20(2)$ & $3377(2)$ & $2.005(1)$ & $%
4.7\times 10^{20}$%
\end{tabular}
$

\bigskip

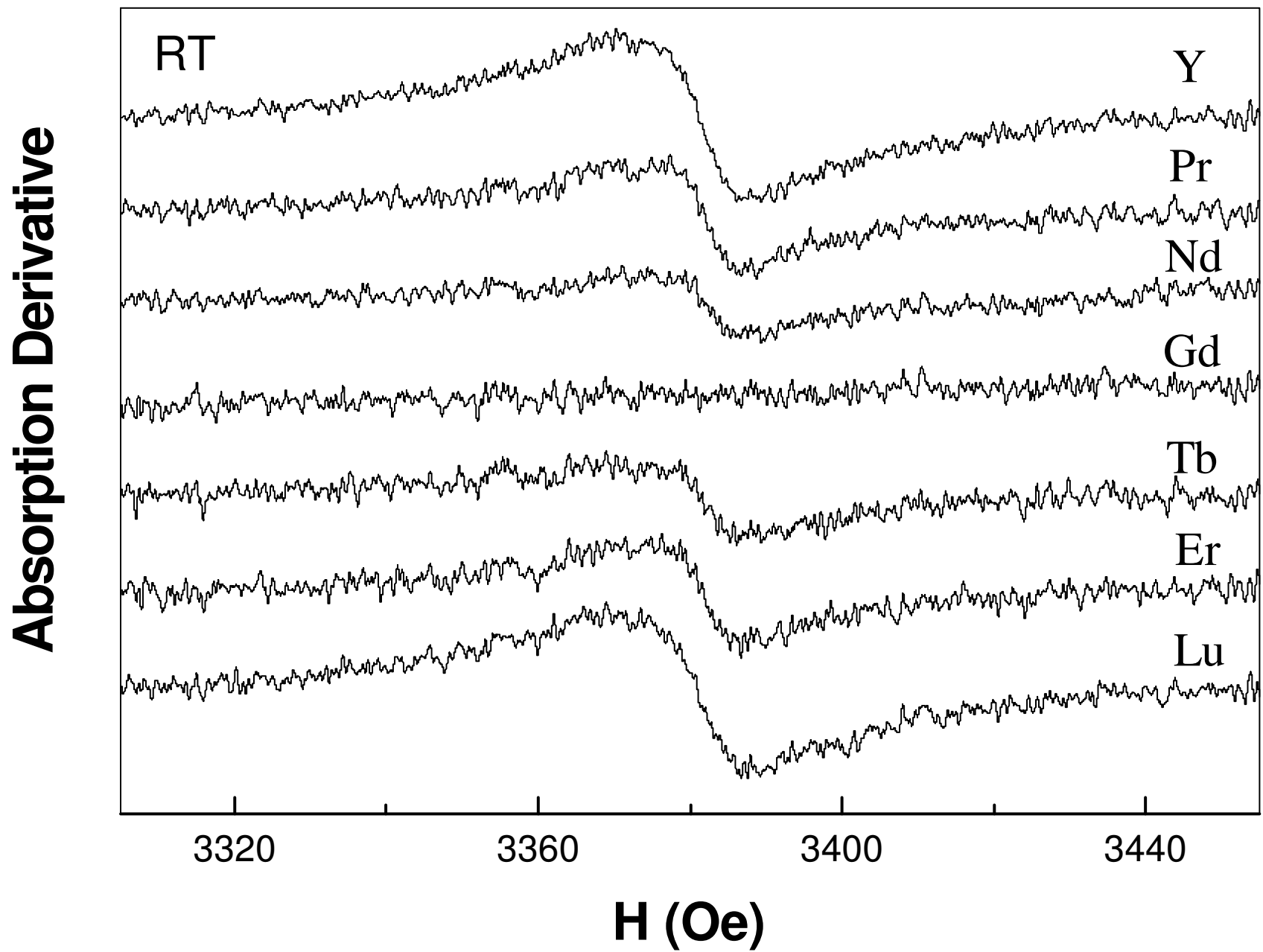
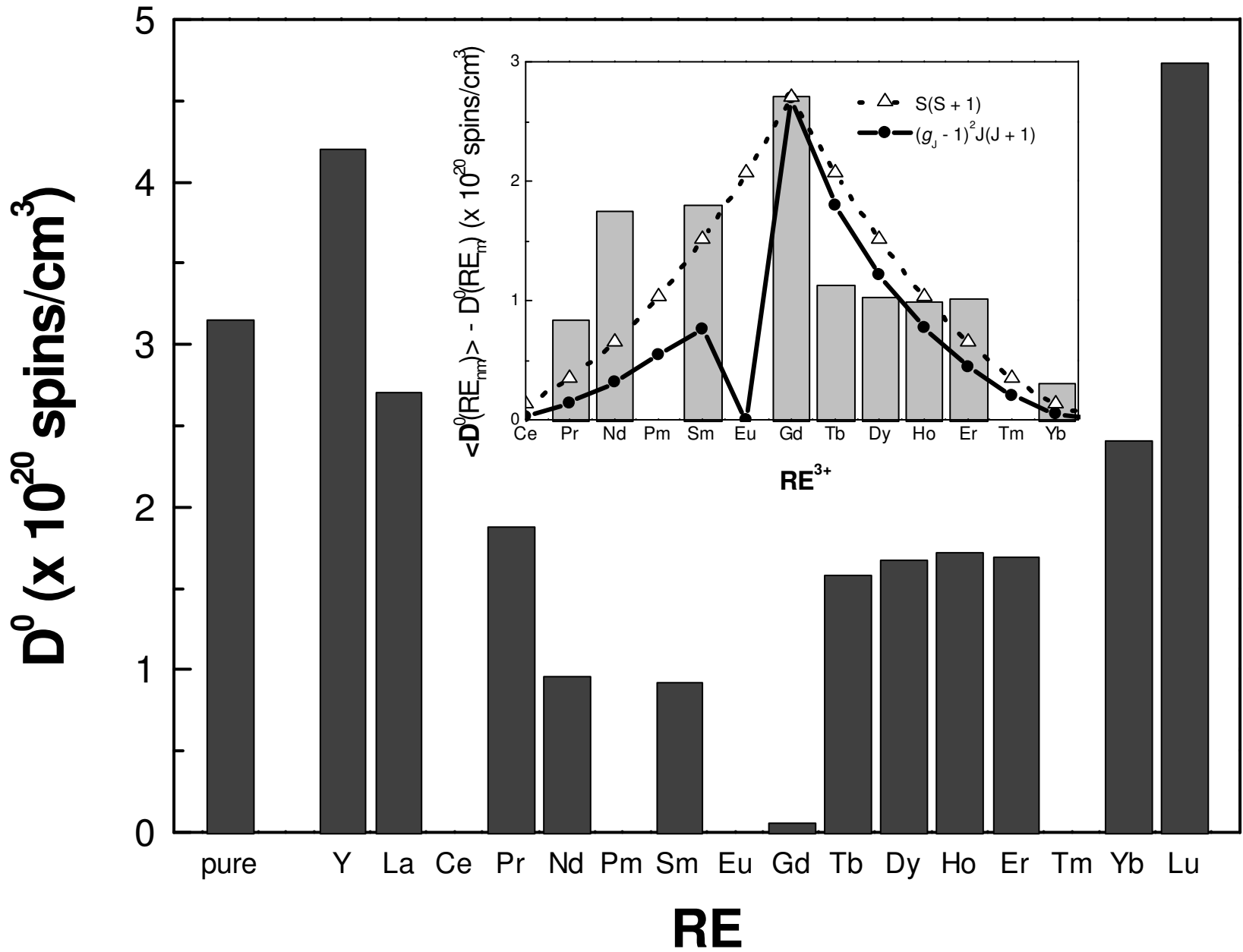
\end{document}